\documentclass{aastex}

\shorttitle{Fern\'andez, Jewitt, and Sheppard}
\shortauthors{Centaurs Asbolus and Chiron}

\begin{document}

\title{Thermal Properties of Centaurs Asbolus and Chiron}

\author{Yanga R. Fern\'andez\altaffilmark{1}, David C. Jewitt\altaffilmark{1},
and Scott S. Sheppard}
\affil{Institute for Astronomy, Univ. of Hawai`i at M\=anoa, \\
2680 Woodlawn Dr., Honolulu, HI 96822}
\email{yan@ifa.hawaii.edu, jewitt@ifa.hawaii.edu, sheppard@ifa.hawaii.edu}

\altaffiltext{1}{Visiting Astronomer at W. M. Keck Observatory, which is 
jointly operated by the California Institute of Technology and the 
University of California.}

\begin{abstract}

We have measured the mid-infrared thermal continua from two Centaurs,
inactive (8405) Asbolus and active 95P=(2060) Chiron, and have
constrained their geometric albedos, $p$, and effective radii, $R$,
with the Standard Thermal Model for slow rotators. These are the first
such measurements of Asbolus; we find $R=33{\rm\ km} \pm 2$ km and
$p=0.12\pm0.03$. This albedo is higher than all of those confidently
known for active cometary nuclei. The thermal inertia is comparable to
or lower than those of main belt asteroids, the Moon, and Chiron; lower
than those of the icy Galilean satellites; and much lower than those of
near-Earth asteroids. For Chiron, we find $R=74{\rm\ km} \pm 4$ km and
$p=0.17\pm0.02$.  While this albedo is consistent with the established
value, previous radiometry by others implied a larger radius. This
effect may be partially due to a varying infrared dust coma but all
datasets have too low signal to be sure. Four Centaur albedos (out of
about 30 objects) are now known.  They show a diversity greater than
that of the active comets, to which they are evolutionarily linked.

\end{abstract}


\section{Introduction}

The so-called Centaur population consists of objects with perihelia
beyond Jupiter but orbital semimajor axes smaller than Neptune's.
Their orbits are unstable on timescales of about $10^{5.5}$ to
$10^{6.5}$ yr \citep{dld96} due to perturbations by the giant planets
and are in dynamical transition from the trans-Neptunian Kuiper Belt to
the inner Solar System. Those Centaurs that are not thrown from the
Solar System or impact planets or the Sun are transferred into typical
Jupiter-family comet-like orbits; thus the Centaurs hold a direct
evolutionary link between the cometary and trans-Neptunian populations.
The colors of many Centaurs, extinct comet candidates, and active
comets \citep{har87,luu93} are consistent with this dynamical picture,
and further physical and compositional studies will let us understand
the physical evolution. Moreover, since the Centaurs are generally
brighter than the trans-Neptunian objects (TNOs), they provide a proxy
through which we can infer the ensemble properties of that more distant
population.

The spectral diversity of the Centaurs and TNOs is well established
\citep[e.g.,][]{lj96,dav98,jl98,blt99}, but the primary cause of this
phenomenon is unknown.  A secondary cause is probably the competition
between reddening by cosmic-ray-induced surface chemistry and impacts
exposing sub-surface icy material \citep{jl01}.  Such a mechanism might
also have observable effects on the albedo as a function of object size
and color. Furthermore, cometary activity, e.g. as shown by Chiron, may
significantly influence the albedo as icy grains on ballistic
trajectories dust the surface.  Thus, sampling the Centaur and TNO
albedos could provide clues to the physical nature of their surfaces.

In this paper we describe the determination of the geometric albedos
and sizes of two Centaurs, Chiron and Asbolus, using the radiometric
method.  We will place the Chiron dataset in the context of earlier
work by \citet{leb84}, \citet{cam94}, \citet{as95}, and \citet{gro01}.

\section{Observations and Reduction}

The observations span two wavelength regimes, mid-infrared (MIR) and
visible. The MIR data were obtained with the Keck I telescope using the
``LWS'' $256^2$-pixel camera \citep{jp93}, and the visible-wavelength
data were obtained with the UH 2.2-m telescope using a Tektronix
$2048^2$-pixel charge-coupled device (CCD). Table 1 gives the
observational circumstances of the two targets and the measured fluxes.
The objects were unresolved at MIR wavelengths.  No visible-wavelength
data were obtained for (8405) Asbolus so for our analysis we use
results published by others.

The MIR data were obtained using chopping and nodding, both with throws
of 4 arcsec. Non-sidereal guiding was used for each target. Flat fields
were obtained by comparing staring images taken at both high and low
airmass.  The seeing was about 0.3 arcsec full-width at half-maximum
(FWHM) at 12.5 $\mu$m and 0.45 arcsec FWHM (diffraction-limited) at
17.9 $\mu$m.  Flux calibration was done by comparing count rates to the
following known (12.5 and 17.9 $\mu$m) flux densities of standard
stars:  $\alpha$ Lyr, 26.4 Jy and 12.9 Jy; $\sigma$ Lib, 120.7 Jy and
58.9 Jy; $\alpha$ CrB, 3.64 Jy and 1.97 Jy; $\gamma$ Aql, 54.2 Jy and
27.5 Jy. These values are derived from the standard system in use at
UKIRT\footnote{This information is available at World Wide Web URL
\hfil\break {\tt
http://www.jach.hawaii.edu/JACpublic/UKIRT/astronomy/conver.html}.}
\citep{chr98} and the magnitudes given by \citet{tok84}.  We corrected
for atmospheric extinction by comparing the stars' photometry over a
range of airmasses.  As the filters we used are only 10\% wide, the
correction to monochromatic magnitudes was 0.01 mag or less, and so was
ignored.

The visible-wavelength image was obtained while guiding on a nearby
star at a sidereal tracking rate in seeing that was 1 arcsec FWHM. A
flat field was constructed by combining dithered images of the blank
twilight sky. Flux calibration and airmass corrections were obtained by
repeated measurements of the \citet{lan92} standard stars SA 107-457,
104-485, and 112-250.  Chiron displayed a faint coma, as was expected,
but the light was dominated by the nucleus. We used an aperture of
radius 1.1 arcsec to minimize the coma's contribution.

\section{Analysis}

\subsection{Thermal Model}

The basic radiometric method to obtain an effective radius, $R$, and
geometric albedo, $p$, is to solve two equations with these two unknowns,
first done about 30 years ago \citep{all70,mat70,mor73} and 
described in detail by \citet{ls89}:
\begin{mathletters}
\begin{eqnarray}
F_{vis}(\lambda_{vis}) & = & 
                {{F_{\odot}(\lambda_{vis})}\over{(r/1{\rm AU})^2}}\ \pi R^2 p\ 
                                 {{\Phi_{vis}}\over{4\pi\Delta^2}}, \\
F_{mir}(\lambda_{mir}) & = & 
        \epsilon\!\int\!B_\nu(T(pq,\theta,\phi),\lambda_{mir}) 
		d\phi d\cos\theta\ R^2\ {{\eta \Phi_{mir}}\over{4\pi\Delta^2}},
\end{eqnarray}
\end{mathletters}
where $F$ is the measured flux density (in e.g. W m$^{-2}$ Hz$^{-1}$)
of the object at wavelength $\lambda$ in the visible (``vis'') or
mid-infrared (``mir''); $F_{\odot}$ is the flux density of the Sun at
Earth as a function of wavelength; $r$ and $\Delta$ are the object's
heliocentric and geocentric distances, respectively; $\Phi$ is the
(dimensionless) phase function in each regime; $B_\nu$ is the Planck
function (in e.g. W m$^{-2}$ Hz$^{-1}$ sr$^{-1}$); $\epsilon$ is the
(dimensionless) infrared emissivity; $\eta$ is a (dimensionless) factor
to account for infrared beaming; and $T$ is the temperature. The
temperature itself is a function of the geometric albedo $p$, surface
planetographic coordinates $\theta$ and $\phi$, and the (dimensionless)
phase integral $q$ which links the geometric and Bond albedos.  For
lack of detailed shape information -- as is the case for our two
objects -- the modeled body is assumed to be spherical.

We employ the ``standard thermal model'' (STM) for slow-rotators
\citep{leb86} to derive the function $T$ and evaluate Eqs. 1. In the
STM, the rotation is assumed to be so slow and/or the thermal inertia
so small that every point on the surface is in instantaneous
equilibrium with the impinging solar radiation. We will show below that
the other extreme, a model assuming a fast-rotator with a rotation axis
perpendicular to the Sun-Earth-object plane, is inconsistent with the
measured color temperatures.

The other parameters to the models are $\epsilon$, $\eta$,
$\Phi_{mir}$, $\Phi_{vis}$, and $q$. Emissivity is close to unity and
we will assume $\epsilon=0.9$ here.  The beaming parameter is known for
only a few of the largest asteroids, but to facilitate comparison with
other work we adopt the standard value $\eta=0.756$ \citep{leb86}.  For
$\Phi_{mir}$ we assume that the magnitude scales with the phase angle
$\alpha$: $-2.5\log\Phi_{mir} = \beta\alpha$, where, based on earlier
work \citep{mat70,leb86}, $0.005$ mag/deg $\le\beta\le0.017$ mag/deg.
In the much-better studied visible regime, we use the $H,G$ formalism
\citep{lb81,rch85} to obtain $\Phi_{vis}$.  The slope parameter $G$
ranges between 0.0 and 0.7. The value of $G$ determines $q$, but since
that has a minor effect on the modeling we adopt $q=0.38$, the
integral's value for $G=0.15$.  Note that our observations all occurred
at small $\alpha$ (Table 1).

We should note that the values for $R$ and $p$ are valid in the context
of the model used, but since the thermal model represents an extremum
of thermal behavior it is not a perfect descriptor. The error bars here
and in many other published reports usually do not describe the
systematic errors from the model itself.  However, such errors are
likely to be comparable to the quoted formal errors so the values are
still physically meaningful.


\subsection{Modeling Results for Asbolus}

We did not obtain our own visible-wavelength measurements of Asbolus,
but photometry by \citet{bl97} provides an excellent constraint. They
measured a mean absolute magnitude\footnote{Absolute magnitude is the
hypothetical apparent magnitude of an object when 1 AU from Earth, 1 AU
from the Sun, and at zero phase angle.} of $H = 8.43 \pm 0.05$ in $R$
band.  Since our MIR measurements were taken at an unknown rotational
phase, and the peak-to-valley amplitude is about 0.4 mag, we adopt a
$\pm0.2$ mag uncertainty. With this visible-wavelength information and
our June 23 mid-IR photometry (see Table 1), the STM provides $T$ and
Eqs. 1 yield Asbolus' effective radius $R$ and geometric albedo $p$.
Since there are not enough data points to perform the $\chi^2$
statistical test, we have found instead the range of values for $R$ and
$p$ such that the model passes within 2$\sigma$ of all data points. The
ranges, means ($\bar R$ and $\bar p$), and standard deviations are:
\begin{mathletters}
\begin{eqnarray}
\bar R = 33{\rm\ km} \pm 2{\rm\ km}; & & 29{\rm\ km} \le R \le 38 {\rm\ km} \\
\bar p = 0.12 \pm 0.03; & & 0.06 \le p \le 0.20
\end{eqnarray}
\end{mathletters}
We note that the object was a few mJy brighter on June 21 compared to
June 23; the time difference is 5.35 rotation periods
\citep{bl97,dav98}, and the magnitude difference is $0.20\pm0.18$.
This is consistent with rotational modulation caused by a changing
cross section, since in that case the mid-IR and visible amplitudes
would be the same. However we do not know the exact rotational phase of
the mid-IR data since the uncertainty in the magnitude difference is
too big. (Otherwise we could find the position on a sine curve
corresponding to that shift in rotation phase and change in
brightness.) At worst our data correspond to a minimum or maximum in
brightness, but a 0.20 mag shift would affect $R$ by only about 1\%
(since the object is dark) and $p$ by 15 to 20\%.

The STM predicts a (12.5- to 17.9-$\mu$m) color temperature $T_c=135$
K, while our photometry shows $T_c=144$ K $\pm13$ K.  By comparison,
the fast-rotator model mentioned previously predicts $T_c=100$ K, so
that model is inapplicable. Since the spin period of Asbolus is not
exceptionally slow -- about 8.9 h \citep{bl97,dav98} -- the other
contributing effect, the thermal inertia, must be low.  A caveat is
that the fast-rotator model degenerates into the slow-rotator model for
a rotation axis pointing at the Sun, which could deceive us in
interpreting the thermal inertia.  However, the fact that (a) Asbolus
has a large photometric amplitude \citep{bl97} and (b) \citet{krn00}
have reported spectroscopic variation over the course of a rotation
make a pole-on point of view unlikely.  Note that the lack of
rotational context prevents us from matching our albedo to the reported
spectroscopic variation.

Given that the object is a slow rotator, we can calculate an upper
limit to the thermal inertia. The dimensionless thermal parameter
$\Theta$, introduced by \citet{sls89}, is defined as
\begin{equation}
\Theta = { {\Gamma \sqrt{\omega}}\over{\epsilon\sigma T_{SS}^3}},
\end{equation}
where $\Gamma$ is the thermal inertia, $\omega$ is the rotational
frequency, $\sigma$ is the Stefan-Boltzmann constant (in e.g.  W
m$^{-2}$ K$^{-4}$), and $T_{SS}$ is the subsolar equilibrium
temperature. This parameter is less than unity for a slow-rotator (and
zero for a body that is non-rotating or has no thermal inertia), and
greater than unity for a fast-rotator.  Thermal inertia itself is
defined as $\Gamma = \sqrt{\kappa\rho c}$, where $\rho$ is the object's
bulk density (in e.g. kg m$^{-3}$), $c$ is the heat capacity (in e.g. J
kg$^{-1}$ K$^{-1}$), and $\kappa$ is the conductivity (in e.g. W
m$^{-1}$ K$^{-1}$).  Thus $\Gamma$ gives clues to the internal thermal
behavior.  At Asbolus' heliocentric distance (with $\epsilon=0.9$,
$p=0.12$, and $q=0.4$), $T_{SS}=142$ K, so $\Theta\le1$ requires that
$\Gamma \le 10.5$ J m$^{-2}$ s$^{-1/2}$ K$^{-1}$.  By comparison, in
the same units, Ceres' value is about 10 \citep{spe90}; the Moon, 50
\citep{ws69}; Europa, about 45-70 \citep{spe99}, Ganymede, 70
\citep{spe87}; near-Earth asteroid Eros, 170 \citep{hd99}; and
near-Earth asteroid Phaethon, $>320$ \citep{hdg98}.  Asbolus' limit of
10.5 argues for a relatively porous and/or rough surface -- perhaps a
regolith, perhaps the aftermath of episodic cometary activity --
inhibiting heat flow. The quantity $\Gamma$ is not yet known for any
typical cometary nucleus.

\subsection{Modeling Results for Chiron}

With our visible-wavelength and 12.5-$\mu$m photometry from Table 1,
the STM provides $T$ and Eqs. 1 yield Chiron's effective radius $R$ and
geometric albedo $p$. Again, since there are not enough data points to
permit the $\chi^2$ statistical test, we have found instead the range
of values for $R$ and $p$ such that the model passes within 2$\sigma$
of all data points. The ranges, means, and standard deviations are:
\begin{mathletters}
\begin{eqnarray}
\bar R = 74{\rm\ km} \pm 4{\rm\ km}; & & 67{\rm\ km} \le R \le 82 {\rm\ km} \\
\bar p = 0.17 \pm 0.02; & & 0.13 \le p \le 0.21
\end{eqnarray}
\end{mathletters}
We note that Chiron appeared slightly brighter on June 21 compared
to June 23 (Table 1), but the magnitude difference, $0.14\pm0.17$, is
consistent with no change. Since Chiron has a small rotational
amplitude, 0.1 mag or less, this is expected and finding the rotational
context is not as critical as for Asbolus.

The STM predicts a (12.5- to 17.9-$\mu$m) color temperature of
$T_c=120$ K. Our photometry shows $T_c=155$ K $\pm20$ K, which is about
1.75$\sigma$ higher but clearly a better match than the fast-rotator
model, which predicts $T_c=88$ K. Since the spin period of Chiron is
not exceptionally slow -- about 5.9178 h \citep{lj90,mb93} -- the
thermal inertia must be low.  The same inertia caveat as for Asbolus
may be applied here, but there is some evidence that we do not view the
axis pole-on. The photometric variation due to rotation, which is less
than 0.1 mag \citep{lj90}, does not change with ecliptic longitude
after accounting for damping by the coma \citep{mb93}.

Independently, \citet{gro01} have used a ``mixed model'' thermal model
of Chiron's surface, integrating the thermal properties of water ice
and refractory grains, to constrain the thermal inertia from {\sl
ISO}+ISOPHOT mid- and far-IR photometry \citep{pes97}. They obtain
$\Gamma=10$ J m$^{-2}$ s$^{-1/2}$ K$^{-1}$, similar to Asbolus'
limiting value from above.

Critical for the determination of the albedo is a robust measurement of
the nucleus' visible-wavelength flux density without contamination from
comatic light. As a check, we can compare our measured magnitude to the
long-term photometric behavior of Chiron.  In early 1985 Chiron appears
to have been at one of its intrinsically faintest points \citep{mb93},
judging by the light curve over the last 30 years \citep{laz97,bus01}.
\citet{mb93} report a V band absolute magnitude $H_v=6.84$, assuming
the slope parameter $G\approx0.7$. The uncertainty in $G$ is high and
introduces several tenths of a magnitude of uncertainty to $H_v$, so
here we shall instead refer magnitudes to the phase angle of their
observations, $\alpha=\alpha_0=3.19^\circ$ on UT 1985 January 19. This
is convenient not only because we avoid having to worry about most of
the opposition surge, but also because Chiron had a similar phase angle
during our visible-wavelength observations.  Thus, whereas $H_v = V -
5\log(r\Delta) + 2.5\log\Phi_{vis}(\alpha)$, let us define
\begin{equation}
H_{v,\alpha_0} = V - 5\log(r\Delta) + 2.5\log\Phi_{vis}(\alpha) - 
     2.5\log\Phi_{vis}(\alpha_0).
\end{equation}
From the data published by \citet{mb93} for UT 1985 January 19, we
calculate that $H_{v,\alpha_0} = 6.96\pm0.01$.

Now, regarding our photometry, the phase angle was
$\alpha=2.70^\circ$.  If $0.0\le G\le0.7$, which is a range that covers
nearly all of the asteroids for which slope parameters are known, then
$2.5\log\Phi_{vis}(\alpha)-2.5\log\Phi_{vis}(\alpha_0) = 0.03\pm0.01$
mag.  Chiron has nearly solar colors \citep{har90}, so $V-R=0.37$, and
substituting the data from Table 1 into Eq. 5 yields $H_{v,\alpha_0} =
6.85\pm0.03$. Thus Chiron was only about 0.11 mag brighter than its
faintest point in early 1985.  Since the rotational phase was unknown,
effectively the difference is $0.11\pm0.05$ mag, but the nucleus
provided about 90\% of the measured flux density during our
observations.

\section{Discussion}
\subsection{Albedo Context}

Figures 1 and 2 display our current understanding of albedos among the
Centaurs and related bodies. The data were taken from this work and
from many other published sources; the plotted error bars are those
cited by the various authors.

In Fig. 1, we plot albedo vs. effective radius (top) and vs. perihelion
distance (bottom). We have only included objects with both reliable
radii and albedos, and we have excluded Pluto since its surface is
strongly influenced by atmospheric effects. A trend with perihelion
would suggest that the albedo is altered by thermal processing from
insolation, but there is no apparent correlation in the plot. The
addition of active and dormant comets in the 2-5 AU range would be
useful.  A trend with radius among the outer Solar System objects might
imply a connection between the albedo and effects that depend on cross
section (such as the impact rate), and here, if we calculate the
linear-correlation coefficient, there is a correlation on the 3$\sigma$
level (as also noted by \citet{jae01}). However this correlation is
solely due to Charon, at $R=625$ km, $p=0.38$.

The distribution of albedos itself however readily reveals that there
is a greater diversity among the Centaurs than among the comets.
Activity might yield a spuriously high albedo -- since in that case one
could overestimate the visible-wavelength flux -- but for the inactive
Asbolus that is inapplicable.  It appears that during the dynamical
cascade from the Kuiper Belt, through the Centaur region, and into the
inner Solar System, an object does not necessarily preserve its
albedo.  Whether this is just a bias in our sampling of
differently-sized objects -- since we have not yet seen many
Centaur-sized, inner Solar System, active comets -- remains to be seen.
However the effect may provide a clue to the mechanisms of cometary
activity, since the phenomenon occurring on Chiron does not appear to
leave behind the same dark, mantled surface as we infer on the cometary
nuclei. For example, an object that erupts in the Centaur region may be
exposing pristine ice and/or dust the surface with icy grains.  This
scenario would suggest that Asbolus may still be sporadically active or
only recently have entered a quiescent state.

Figure 2 compares albedo with $B-V$, $V-R$, and $V-J$ colors.  Again
the population is sparse but we note a general trend of redder objects
having a lower albedo.  This would be consistent with the idea that
chemistry activated by cosmic ray irradiation effects surface darkening
and reddening -- if there were a mechanism to brighten the surface in
the first place. For the Centaurs, cometary activity could be that
mechanism. Impacts contribute to the effect as well but since most
impactors are too small to cause widespread resurfacing, and since
other color data now are inconsistent with the predicted
manifestations, this cannot be the dominant cause \citep{jl01}. The
full explanation of the color diversity remains unknown, but in any
case populating Fig. 2 with the albedos of more ordinary objects would
be wise.

\subsection{Earlier Chiron Radiometry}

Table 2 gives a list of published radiometric estimates of Chiron's
radius (with 1$\sigma$ uncertainties) and color temperature as a
function of time and heliocentric distance.  The average radius
(weighted by the variances) is 82 km; on the 2$\sigma$ level all points
are consistent. Some of the variation on the positive side would
normally be attributed to contamination from the infrared dust coma,
but this would not mimic the long-term trend in activity seen at
visible wavelengths \citep{laz97}.  Moreover a high color temperature
is difficult to explain.  Grains much smaller than the wavelength of
peak Planck emission will appear to be hotter than expected but if that
were the case with Chiron one would also expect the comet to be
intrinsically brighter (and hence yield a larger effective radius),
opposite to what is observed. A further constraint is the occultation
result of \citet{bus96}, where the assumed-spherical Chiron needs to
have $R\ge90$ km. Since rotational light curves imply that Chiron is
indeed only about 10\% aspherical \citep{bus89}, this would imply a
minimum effective radius of about 85 km.

The reconciliation of the radiometry in Table 2 (and the occultation
data) may involve more detailed thermal modeling, as \citet{gro01} have
done, but further observations are certainly necessary since none of
the thermal measurements in the past 20 years have high
signal-to-noise. Thus it is possible that photometric uncertainty is
corrupting our interpretation.

Lastly, we note that the albedo in Eq. 4b is consistent with previous
values \citep{cam94}. Since our visible photometry is about 0.1 mag
brighter than the faintest intrinsic measurements, as discussed in \S
3.3, such a correction would reduce our albedo by 0.015. \citet{gro01}
report an albedo of $0.12\pm0.02$ using $H_v = 6.95\pm0.2$, which would
make Chiron about 0.3 mag intrinsically fainter than observed by
\citet{mb93} in 1985.  A correction of that size would further reduce
our albedo by 0.037.  Thus all variation among reported albedos can be
explained by the use of the different absolute magnitudes.

\section{Summary}

1. We have radiometrically determined the effective radii $R$ and 
geometric albedos $p$ of Centaurs Asbolus and Chiron:
\begin{displaymath}
29{\rm\ km} \le R \le 38 {\rm\ km,\ and\ } 0.06 \le p \le 0.20
                                        {\rm,\ for\ Asbolus.}
\end{displaymath}
\begin{displaymath}
67{\rm\ km} \le R \le 82 {\rm\ km,\ and\ } 0.13 \le p \le 0.21
                                       {\rm,\ for\ Chiron.}
\end{displaymath}
The ranges effectively cover 2$\sigma$.

2. Under the STM formalism \citep{leb86,sls89}, we calculate an upper
limit to Asbolus' thermal inertia is $\Gamma \le 10.5$ J m$^{-2}$
s$^{-1/2}$ K$^{-1}$. This value is comparable to those of main belt
asteroids, smaller than those of icy Galilean satellites, and an order
of magnitude smaller than the few values known for near-Earth
asteroids. \citet{gro01} find a comparable value for the $\Gamma$ of
Chiron. Presumably the surfaces of Chiron and Asbolus are relatively
porous and/or rough enough to inhibit heat flow.

3. Chiron's albedo is consistent with earlier measurements by others,
taking into account the varying absolute magnitude, but others have
noted a larger radius (at the 2$\sigma$ level). Whether this is due to
a varying infrared coma or simply the low signal-to-noise of all extant
radiometric data remains to be seen.

4. Of the four Centaur albedos now known, two are comet-like and two
are 2 to 3 times higher.  During the dynamical cascade of objects from
the Kuiper Belt and through the Centaur region into the inner Solar
System, the albedo is apparently not always preserved, although a more
robust demonstration would require a comparison of similarly-sized
objects. Nevertheless, there is a greater diversity among the albedos
of the Centaurs than of the cometary nuclei.

\acknowledgments

We thank Jane Luu for helpful comments on the manuscript.  We
gratefully acknowledge the help of Keck staff members Joel Aycock, Meg
Whittle, Wayne Wack, and Greg Wirth; and 2.2-m staff member John
Dvorak. We also acknowledge the JPL Solar System Dynamics group for
their Horizons on-line ephemeris generation program. This work was
supported by grants to DCJ from NSF.


\clearpage

\figcaption[]{Plot of well-determined albedos vs. well-determined radii
(top) and vs. perihelion distance (bottom). We have included only
active comets (red squares), Centaurs (blue circles), and
trans-Neptunian objects (TNOs; green diamonds). For clarity of the
other points, Charon (0.38 albedo) is off the top of each plot.  Error
bars are reported 1$\sigma$. With four objects, the Centaurs already
have a greater diversity in albedo than the known cometary nuclei.  The
plotted data were obtained from this work,
\citet{ahe89},  \citet{cam87},  \citet{cam94},  \citet{cam95},  
\citet{dav93},  \citet{fer99},  \citet{fer01},  \citet{han85},  
\citet{jae01},  \citet{jk98},   \citet{jor00},  \citet{kel86},  
\citet{mac88},  and \citet{tho00}.}  

\figcaption[]{Plot of well-determined albedos and known $B-V$, $V-R$,
and $R-J$ colors. Symbol shapes are the same as for Figure 1.  There is
a general trend of redder objects being darker, although the sampling
will have to be improved before there is high confidence in this
conclusion, since many of the objects in the plot are atypical. The
color data were obtained from
\citet{bin88}, \citet{bl97}, \citet{dav98} \citet{har90}, 
\citet{jl98},  \citet{jl01}, \citet{jew01}, \citet{lj96}, 
\citet{rom97}, and \citet{syk00}. We have used the Pluto-Charon systemic 
color for $V-R$ and $R-J$.}


\clearpage
\begin{table}
\begin{center}
\caption{Observing Circumstances \vskip 5pt}
\scriptsize
\begin{tabular}{cccccccccc}
\tableline\tableline
Object & Date & Time & Exp. & Airmass & $\lambda$ & 
					$r$ & $\Delta$ & $\alpha$ & 
								Flux Density\\
       & (UT) & (UT) & (s)  &         & ($\mu$m) &
                                        (AU) & (AU)    & (deg)  &  \\
\tableline
(2060) Chiron  & 2000 Jun 21 & 09:31 & 324 & 1.25 & 12.5 & 10.109 & 9.139 & 1.79  & $14.4\pm1.8$ mJy \\
      ''       & 2000 Jun 21 & 10:02 & 324 & 1.28 & 12.5 &    ''  &   ''  &  ''   & $18.2\pm2.7$ mJy \\
      ''       & 2000 Jun 23 & 10:42 & 324 & 1.39 & 12.5 & 10.113 & 9.153 & 1.98  & $14.3\pm1.7$ mJy \\
      ''       & 2000 Jun 23 & 10:58 & 309 & 1.45 & 17.9 &    ''  &   ''  &  ''   & $43.9\pm15.9$ mJy \\
      ''       & 2000 Jun 23 & 11:14 & 309 & 1.52 & 17.9 &    ''  &   ''  &  ''   & $56.4\pm16.8$ mJy \\
      ''       & 2000 Jul 01 & 09:54 & 120 & 1.35 & 0.65 & 10.127 & 9.218 & 2.70  & $16.30\pm0.01$ mag\tablenotemark{a}  \\
(8405) Asbolus & 2000 Jun 21 & 08:54 & 324 & 1.91 & 12.5 &  7.867 & 6.988 & 3.94  & $17.0\pm3.0$ mJy \\
      ''       & 2000 Jun 21 & 09:11 & 324 & 1.96 & 12.5 &    ''  &   ''  &  ''   & $19.7\pm2.7$ mJy \\
      ''       & 2000 Jun 23 & 08:40 & 324 & 1.90 & 12.5 &  7.862 & 6.998 & 4.13  & $15.2\pm1.9$ mJy \\
      ''       & 2000 Jun 23 & 08:57 & 309 & 1.94 & 17.9 &    ''  &   ''  &  ''   & $50.4\pm15.7$ mJy \\
\tableline
\end{tabular}
\tablenotetext{a}{Object was not a point-source. We used a circular synthetic 
aperture of radius 1.1 arcsec.}
\tablecomments{``Exp.'' gives the on-source integration time. The wavelength 
of observation is $\lambda$. The heliocentric and geocentric distances are 
$r$  and $\Delta$, respectively. The phase angle is $\alpha$.}
\end{center}
\end{table}

\clearpage
\begin{table}
\begin{center}
\caption{Chiron Radiometric Radii \vskip 5pt}
\scriptsize
\begin{tabular}{ccccccl}
\tableline\tableline
UT Day & $r$ (AU) & $R$ (km) & $T_c$ (K) & $T_e$ (K) & Ref. & Notes \\
\tableline
1983 Jan 9/10  & 15.8 & $77^{+15}_{-19}$ &   ---      & --- & 1 &  a \\
1991 Nov 18/19 & 10.0 & $74\pm11$        & $143\pm19$ & 122 & 2 &  b \\
1993 Nov 22    &  8.9 & $96\pm7$         & $122\pm10$ & 129 & 2 &  c \\
1994 Mar 31    &  8.8 & $93\pm5$         & $136\pm12$ & 129 & 2 &  c \\
1994 Apr 27    &  8.8 & $84\pm10$        &   ---      & --- & 3 &  a \\
1996 Jun  8-15 &  8.5 & $80^{+3}_{-10}$  & $115\pm7$  & 127 & 4 &  d \\
2000 Jun 21/23 & 10.1 & $74\pm4$         & $155\pm20$ & 121 & 5 &  b \\ 
\tableline
\end{tabular}
\tablenotetext{a}{Only one wavelength was measured, so color not applicable.}
\tablenotetext{b}{Reported $R$ is from just one of the wavelengths.}
\tablenotetext{c}{Listed $R$ is weighted average of the reported results.}
\tablenotetext{d}{Used 25 and 60 $\mu$m data to calculate $T_c$ and $T_e$.}
\tablecomments{Heliocentric distance is $r$, effective radius is $R$,
measured color temperature is $T_c$, and the expected color temperature
from the thermal model is $T_e$. Perihelion occured on 1996 February 15.}
\tablerefs{(1) \citealt{leb84}; (2) \citealt{cam94}; (3) \citealt{as95};
(4) \citealt{gro01};
(5) this work.}
\end{center}
\end{table}

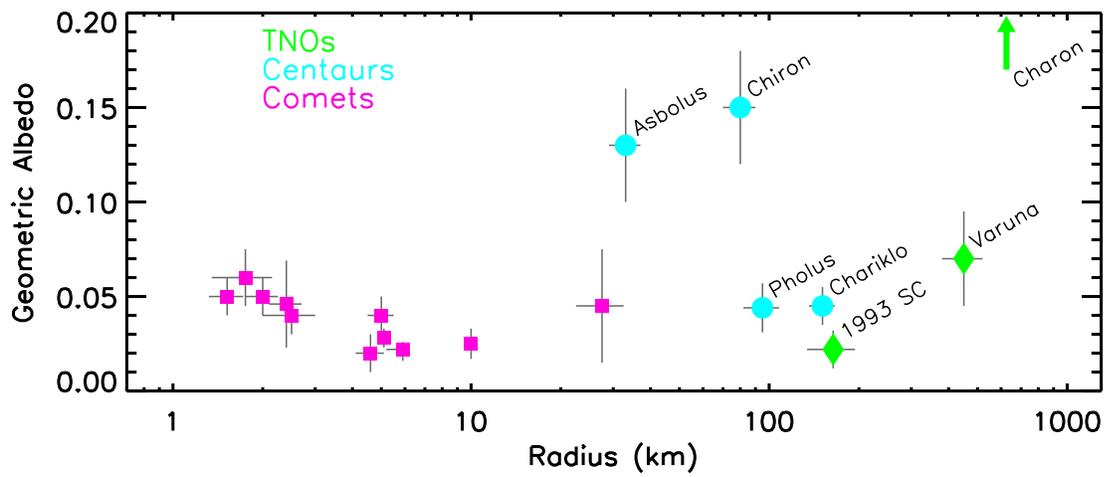
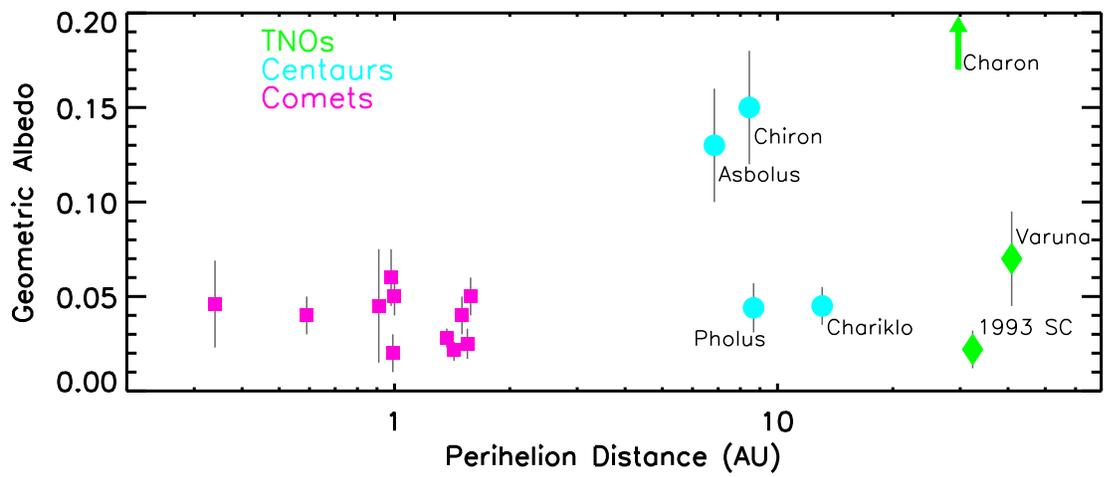
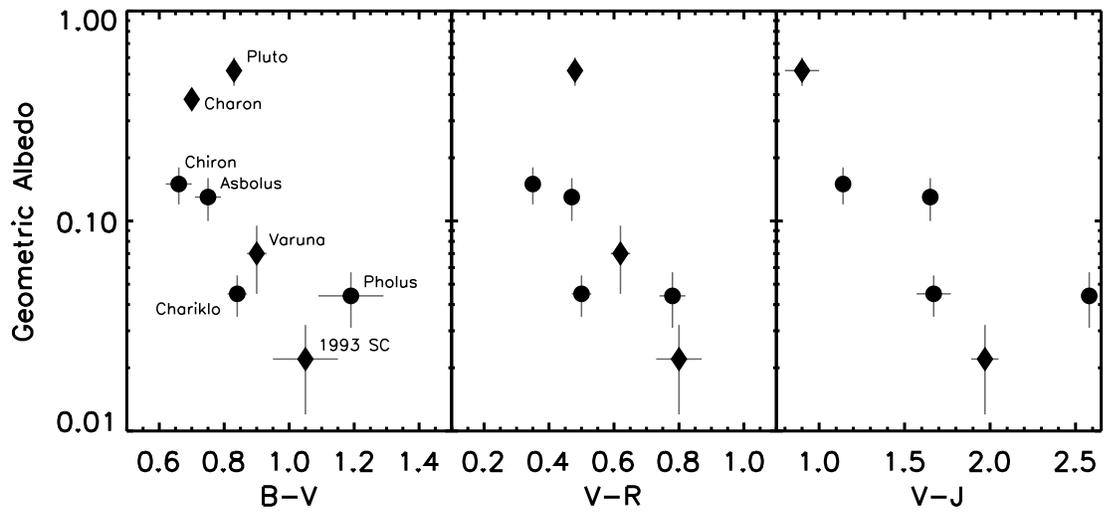
\end{document}